# X-Ray diffraction studies on asymmetrically broadened peaks of heavily deformed Zirconium based alloys

A. Sarkar, P. Mukherjee[*] and P. Barat

Variable Energy Cyclotron Centre

1/AF Bidhan Nagar, Kolkata 700 064, India

## Abstract

The diffraction peaks of Zircaloy-2 and Zr-2.5%Nb alloys at various deformations are found to be asymmetric in nature. In order to characterize the microstructure from these asymmetric peaks of these deformed alloys, X-Ray Diffraction Line Profile Analysis like Williamson-Hall technique, Variance method based on second and fourth order restricted moments and Stephens model based on anisotropic strain distribution have been adopted. The domain size and dislocation density have been evaluated as a function of deformation for both these alloys. These techniques are useful where the dislocation structure is highly inhomogeneous inside the matrix causing asymmetry in the line profile, particularly for deformed polycrystalline materials.

## 1. Introduction

One of the most powerful methods of determination of the dislocation structure in the heavily cold worked materials is the analysis of the broadening in the X-ray diffraction lines. Wilkens [1,2] first developed the theory for symmetrical X-ray diffraction lines broadened by dislocation, which was successfully verified by the experimental study of tensile deformed Cu- single crystals oriented for single slip [3,4].

[*] Corresponding author, paramita@veccal.ernet.in

In case of heavily deformed polycrystalline materials with high Stacking Fault Energy (SFE), the dislocations entangle to form cell structure causing heterogeneity in the dislocation arrangement [5,6]. In the dislocation cell structure, the dislocation cell walls are encapsulating the cell interiors in which the dislocation density can be one or two orders of magnitude smaller than in the cell walls [5-7]. The residual long-range internal stresses are attributed to the interfacial dislocations lined up along the interfaces between cell walls and the cell interior. These interfacial dislocations can have either the configuration of small angle grain boundaries or edge dislocations lying in the interfaces, lined up with same burgers vectors [7]. In both cases, the distributions of dislocations are highly inhomogeneous.

The first theory for the broadening of the diffraction peaks was proposed by Warren & Averbach [8], in which it was assumed that the broadening is caused due to size effect and strain effect. Krivoglaz et. al. [9,10] then pointed out that if the strain is only due to dislocation, Warren-Averbach method can not be applied. Considering that the distributions of dislocations are completely random, they worked out an analytical expression for the broadening. But the serious shortcoming of their work was that the Fourier coefficients diverge logarithmically when the crystal size tends to infinity [11]. This problem was solved by Wilkens [12-14] by the introduction of the concept of restricted random dislocation distribution. But Wilken's theory assumed symmetrical line profiles in contrast to the experimental evidence [15]. Although, asymmetric broadening can be understood by combining Wilken's theory with the concept of quasi composite model [15], it is based on a particular dislocation model which we need to assume *a priori*. Groma [11] has developed a variance method which is based on the asymptotic behaviour of the second and fourth order restricted moments. Though the mathematical foundation of this theory is similar to the model published earlier [16,17], it is based only on the analytical properties of the displacement field of straight dislocations and no assumption is made on the actual form of the dislocation distribution and thus it can be employed for the inhomogeneous dislocation distribution.

For deformed polycrystalline materials, dislocations form network of cellular structure at higher deformation. As a consequence, the characteristic peaks observed in XRD pattern of heavily cold worked material become asymmetric in nature. In our earlier work [18], the microstructural parameters of heavily cold rolled Zircaloy-2 and Zr-2.5%Nb alloy have been characterized by X-Ray Diffraction Line Profile Analysis (XRDLPA) using different techniques like Williamson-Hall plot, Integral Breadth method and Modified Rietveld method. The density of dislocation was estimated from the values of the surface weighted average domain size and average microstrain values, considering the dislocation distribution to be random [18]. In this work, we have used the variance method [11, 19] to characterize the microstructure of the same alloys from a different approach. The mathematical formalism of this method is based only on the analytical properties of the displacement field of dislocations. As a result, it can be applied for the inhomogeneous dislocation distributions. Moreover, the advantage of this method is related to the fact that during integration, the statistical error in the measurement may get cancelled out. We have also evaluated the anisotropic distribution of strain in the three-dimensional space for all the samples, which have arisen due to the inhomogeneous dislocation structure during deformation.

## 2. Experimental procedure

Samples of Zircaloy-2 and Zr-2.5%Nb were collected from Nuclear Fuel Complex, Hyderabad, India and annealed at 1023 K for 10 hours in sealed quartz tubes under vacuum and cooled at the rate of 278K per hour. These samples were then cold-rolled at 30%, 50% and 70% deformation for Zircaloy-2 and 30%, 50% and 60% deformation for Zr-2.5%Nb. X-ray Diffraction (XRD) profile for each sample was recorded from the rolled surface (polished) using PHILIPS 1710 X-ray diffractometer using $CuK_\alpha$ radiation. The diffraction profiles of these samples at various deformed stages were obtained by varying $2\theta$ from 25° to 100° with a step scan of 0.02°. The time spent for collecting data per step was 4 seconds.

## 3. Methods of Analysis

The characterization of the dislocation microstructure evolved during plastic deformation of polycrystalline material is always a challenging problem in the field of materials science. Microstructural studies of deformed polycrystalline sample from XRDLPA have been a topic of renewed interest for the last two decades. Many new methods have been proposed to extract microstructural information from the XRD line profile. XRD gives information of the bulk properties of a powder or a polycrystalline solid, averaged over the sample. We have adopted the following techniques of line profile analysis to obtain microstructural information from the asymmetrically broadened diffraction profiles.

*Williamson-Hall Technique*

Williamson and Hall plot is a classical method to obtain qualitative information of anisotropy in broadening. Williamson and Hall [20] assumed that both size and strain broadened profiles are Lorentzian. Based on this assumption, a mathematical relation was established between the integral breadth ($\beta$), volume weighted average domain size ($D_v$) and the microstrain ($\varepsilon$) as follows.

$$\frac{\beta \cos\theta}{\lambda} = \frac{1}{D_v} + 2\varepsilon\left(\frac{2\sin\theta}{\lambda}\right) \qquad (1)$$

The plot of $\left(\frac{\beta \cos\theta}{\lambda}\right)$ versus $S = \left(\frac{2\sin\theta}{\lambda}\right)$ gives the value of the microstrain from the slope and domain size from the ordinate intercept. If the points in the WH plot are scattered, i.e., if $\left(\frac{\beta \cos\theta}{\lambda}\right)$ is not a monotonous function of $\left(\frac{2\sin\theta}{\lambda}\right)$, the broadening is termed as anisotropic.

*Variance Method*

The variance method has been applied to determine the domain size and dislocation density from the asymmetric intensity profile. The method is based on the analysis of the moments of the intensity profile. Wilson [21] and Groma [11] have shown that the particle size and strain induced broadening create different asymptotic behavior of the k$^{th}$ order restricted moments defined as:

$$M_k(q') = \frac{\int_{-q'}^{q'} q^k I(q) dq}{\int_{-\infty}^{\infty} I(q) dq} \qquad (2)$$

in which $I(q)$ is the intensity distribution as a function of $q = \frac{2}{\lambda}[\sin(\theta) - \sin(\theta_0)]$, where $\lambda$ is the wavelength of the X-ray, $\theta$ is the diffraction angle and $\theta_0$ is the Bragg angle. Accordingly, the second and fourth order restricted moments have the following asymptotic forms:

$$M_2(q) = \frac{1}{\pi^2 \varepsilon_F} q - \frac{L}{4\pi^2 K^2 \varepsilon_F^2} + \frac{\Lambda \langle \rho \rangle \ln(q/q_0)}{2\pi^2} \qquad (3)$$

and $\dfrac{M_4(q)}{q^2} = \dfrac{1}{3\pi^2 \varepsilon_F} q + \dfrac{\Lambda \langle \rho \rangle}{4\pi^2} + \dfrac{3\Lambda^2 \langle \rho^2 \rangle}{4\pi^2 q^2} \ln^2(q/q_1) \qquad (4)$

where K is the Scherrer constant, L is the so called taper parameter depending on the rate of decrease of the cross sectional area of the crystallites. $\varepsilon_F$ is the average column length or area weighted domain size measured in the direction of the diffraction vector. $\langle \rho \rangle$ is the average dislocation density and $\langle \rho^2 \rangle$ is the average of the square of the dislocation density. $q_0$ and $q_1$ are fitting parameters not interpreted physically. $\Lambda$ is a geometrical constant describing the strength of dislocation contrast and its value is of the order of one. It is seen that the leading two terms of $M_2(q)$ and the first term of $M_4(q)$ originate from the finite domain size ($\varepsilon_F$). The third term of $M_2(q)$ and the second term of $M_4(q)$ are due to density of dislocations ($\langle \rho \rangle$). Finally, the last term of $M_4(q)$ corresponds to average of the square of the dislocation density ($\langle \rho^2 \rangle$).

Borbely and Groma [19] have shown that if the particle size broadening is negligible the evaluated values of average dislocation density obtained from the second and fourth order restricted moments agree well. But, if the particle size broadening is significant, it is not possible to evaluate dislocation density from $M_2$ with sufficient precision. In that case, the dislocation density is evaluated from $M_4$ with sufficient accuracy. It is still required to determine both $M_2$ and $M_4$ restricted moments so that the right background level can be chosen in order to obtain same particle size in both cases.

*Analysis of anisotropic peak broadening (Stephens Model)*

The anisotropic line broadening is frequently observed in powder diffraction pattern and creates serious difficulty in the line profile analysis. Anisotropic strain distribution has been introduced by several authors [22] to model the anisotropic peak broadening. P. W. Stephens [23] proposed a phenomenological model of anisotropic broadening in powder diffraction considering the distribution of lattice metric parameters within the sample. In this model each crystallite is regarded as having its own lattice parameters, with multidimensional distribution throughout the powder sample. The width of each reflection can be expressed in terms of moments of this distribution, which leads naturally to parameters that can be varied to achieve optimal fits.

Let $d^*_{hkl}$ be the inverse of the *d* spacing of the (*hkl*) reflection. Then $d^{*2}$ is bilinear in the Miller indices and so can be expanded in terms related to the covariances of the distribution of the lattice metrics. This leads to an expression in which the variance of $d^{*2}$ is a sum of 15 different combinations of Miller indices in the fourth order. Imposing the symmetry of the hexagonal lattice reduces the number of independent terms to the following three:

$$S^2 = S_{400}(h^4 + k^4 + h^3k + hk^3 + h^2k^2) + S_{004}l^4 + 3.S_{202}(h^2l^2 + k^2l^2 + hkl^2) \quad (5)$$

The anisotropic strain contribution to the angular width in 2θ of the reflection is given by

$$\delta 2\theta = (360/\pi)(\delta d/d)\tan\theta, \quad (6)$$

where $\delta d/d = \pi(S^2)^{1/2}/18000 d^*_{hkl} \quad (7)$

## 4. Results and discussions

*Williamson-Hall technique*

Fig. 1 shows the WH plot for Zircaloy 2 and Zr-2.5%Nb samples at different deformation. From these figures, it is seen that the line broadening is not a monotonous function of the diffraction angle indicating the anisotropic broadening of the line profile in all the cases.

*Variance Method*

Variance method is based on the individual peak analysis. We have applied variance method on the important peaks like (002), (102) and (103) diffraction peaks for the smple of Zircaloy-2 at 30%, 50%, 70% deformation and also for Zr-2.5%Nb at 30%, 50%, 60% deformation. (101) peak which is the highest intensity peak for Zr-based alloys is not considered in this analysis as it showed texturing effect with increasing deformation. Fig. 2 represents the (002) Bragg peaks recorded from Zr-2.5%Nb alloy solid samples at different deformations. Fig. 3 and Fig. 4 show the typical second-order and fourth order moments for the (002) peaks of the Zr-2.5%Nb samples at different deformations. The asymptotic regions of the curves are fitted with equations (3) and (4) respectively. The nature of $M_2$ and $M_4$ suggest that both size and strain broadening are present [19]. The calculations of $M_2$ and $M_4$ have been performed for different level of background. Finally, the background values were chosen in such a way that the calculated M2 and M4 from those data range yielded the same domain size values. The dislocation density is calculated from $M_4$. The area weighted domain size and dislocation density obtained from the fit for different sample are listed in Table 1. The maximum error in the size values is ±10% and for dislocation density the maximum error is ±20%.

Zircaloy-2 is a single α- phase alloy. In the early stages of deformation, the predominant slip occurs on the primary slip planes, which are the primary glide planes and the dislocations form coplanar arrays. As the deformation proceeds, cross slip takes place and multiplication process operates. The cold worked structure forms regions of high dislocation density or tangles, which soon develop into tangled networks or cells. Thus the characteristic microstructure in the cold-worked state is a cellular substructure in which high density dislocation tangles form the cell walls. We found that during the cold deformation of Zircaloy-2, the average size of the cells or the domains did not change significantly from 30% to 70% deformation (Table 1).

Zr-2.5% Nb is a two phase (α-β) alloy. β phase being finely dispersed throughout the matrix, plays an important role in the deformation. The value of $\varepsilon_F$ decreased significantly with increasing deformation for this alloy (Table1). The dislocation generated during deformation of the soft phase β formed loops around the hard phase α and created the dislocation cell structures or the domains. This phenomenon occurred as the generation of dislocations retained the continuity between the two phases, which was necessary to avoid any void and micro-cracks. Thus, the size of the domains or cells for Zr-2.5%Nb decreased with progressive deformation.

*Stephens model*

The mathematical formalism in variance method is based on the analytical properties of strain field of dislocation regardless of the actual form of dislocation distribution. The anisotropic strain field resulting due to dislocation has been modeled in terms of dislocation density but the strain parameters can not be calculated directly. The restricted second-order and fourth order moments are correlated with the domain size and the dislocation density for a particular diffraction peak as shown in equation (3) and (4). So, to account for the anisotropic broadening of the sample, as also seen from the WH plots (Fig. 1), we have adopted the Stephens model for anisotropic peak broadening. This model gives the anisotropic strain distribution in the three-dimensional space which is not obtained from the variance method. Using both the techniques on the deformed polycrystalline sample give an overall information of the inhomogeneous dislocation structure developed during deformation. The Rietveld refinement package GSAS [24] has implemented Stephens's model to account strong anisotropy in the half widths of reflections. We have used GSAS to calculate Rietveld refinement of the line profiles of the Zircaloy-2 and Zr-2.5%Nb alloy at different deformation conditions. The profiles have been fitted without and with using the Stephens model. It is found that the incorporation of Stephens's model improved the quality of the fit. As a typical example, for Zircaloy-2 sample at 50% deformation the fitting parameter is $R_{wp}$=3.65%. Without Stephen's model the fitting parameter was $R_{wp}$=5.36%. This suggests that the Stephen's model fit the data very well.

We have used $S_{hkl}$ as the free parameters to obtain the best fit between the model and the experiment. Since the anisotropic broadening has both Gaussian and Lorentzian components, the entire diffraction pattern has been fitted using a mathematical function which includes both the components.

The graphical representation of the three-dimensional strain distribution is obtained using the refined values of the $S_{hkl}$. The three-dimensional strain distribution plot for Zircaloy-2 and Zr-2.5%Nb at various deformations are shown in Fig. 5 and Fig. 6 respectively. It is seen that for both the samples the strain field is anisotropic at all stages of deformation. This may be attributed to the heterogeneous arrangement of dislocation in the matrix, resulting from the cellular structure developed during plastic deformation.

Thus Stephens's model can well predict the nature of the strain field for deformed polycrystalline materials.

## 5. Conclusions

The asymmetric broadening in the peaks of heavily deformed Zircaloy-2 and Zr-2.5%Nb can be well characterised by the variance method, in terms of domain size and dislocation density. Since the mathematical formalism of this method is based only on the displacement field of dislocation, this technique can be applied to determine the average dislocation density of any deformed polycrystalline material. It is found that the domain size has decreased significantly for Zr-2.5%Nb with increasing the degree of deformation as compared to Zircaloy-2. The values of the average dislocation density in Zr-2.5%Nb was also found to be higher than Zircaloy-2 with increasing deformation. WH plot generated for all the deformed alloys showed that the broadening is highly anisotropic. Stephens model gave a clear picture of the anisotropic strain distribution in three-dimensional space for these samples as a function of deformation.

# References

[1] M. Wilkens: In Fundamental Aspects of Dislocation Theory, edited by J. A. Simmons, R. De. Wit and R. Bollough, vol. II, pp. 1195-1221. Natl. Bur. Stand. (US) Spec. Publ. No. 317. Washington, DC, USA.

[2] M. Wilkens: Phys. Status. Solidi A, 1970, vol. 2, pp. 359-370.

[3]M. Wilkens, K. Herz, H. Mughrabi: Z. Metallkd., 1980, vol. 71, pp. 376-384.

[4] T. Ungar, H. Mughrabi, M. Wilkens: Acta Metall., 1982, vol. 30, pp. 1861-1867.

[5] G. E. Dieter: Mechanical Metallurgy, 1986, (McGraw-Hill, New York).

[6] D. A. Hughes, N. Hansen: Acta Mater., 2000, vol. 48, pp. 2985-3004.

[7] T. Ungar: Defect and Microstructure Analysis by Diffraction, Eds. R. L. Snyder, J. Fiala, H. J. Bunge, 1999, pp. 165-199.

[8] B. E. Warren, B. L. Averbach: J. Appl. Phys., 1952, vol. 23, pp. 1059-1063.

[9] M. A. Krivoglaz, K. P. Ryboshapka: Phys. Met. Metallogr., 1963, vol. 15, pp. 14.

[10] M. A. Krivoglaz: Theory of X-ray and Thermal Neutron Scattering by real Crystals (Plenum Press, New York, 1969), p. 258.

[11] I. Groma: Phys. Rev. B, 1998, vol. 57, pp. 7535-7542.

[12] M. Wilkens: Phys. Status. Solidi, 1962, vol. 2, pp. 807.

[13] M. Wilkens: Acta Metall., 1969, vol. 17, pp. 1155.

[14] M. Wilkens: Phys. Status Solidi A, 1987, vol. 104, pp. 344.

[15] T. Ungar, H. Mughrabi, D. Ronnpagel, M. Wilkens: Acta Metall., 1984, vol. 32, pp. 333.

[16] I. Groma, T. Ungar, M. wilkens: J. Appl. Crystallogr., 1988, vol. 21, pp. 47.

[17] I. Groma, T. Ungar, M. wilkens: J. Appl. Crystallogr., 1989, vol. 22, pp. 26.

[18] P. Mukherjee, A. Sarkar, P. Barat, S. K. Bandyopadhyay, P. Sen, S. K. Chattopadhyay, P. Chatterjee, S. K. Chatterjee, M. K. Mitra: Acta Mater., 2004, vol. 52, pp. 5687-5696.

[19] A. Borbely, I. Groma: Appl. Phys. Lett., 2001, vol. 79, pp. 1772-1774.

[20] G. K. Williamson, W. H. Hall: Acta Metall., 1953, vol. 1, pp. 22.

[21] A. J. C. Wilson: Proc. Phys. Soc., 1962, vol. 80, pp. 286.

[22] P. Thomson, D. Cox and J. B. Hastings: J. Appl. Cryst., 1987, vol. 20, pp. 79-83.

[23] P. W. Stephens: J. Appl. Cryst. 1999, vol. 32, pp. 281-289.

[24] A.C. Larson and R.B. Von Dreele: "General Structure Analysis System (GSAS)", Los Alamos National Laboratory Report LAUR 86-748 (2000).

Table 1: Area weighted domain size and average dislocation density obtained from the second and fourth order restricted moments of the (002), (102) and (103) peaks of Zircaloy-2 and Zr-2.5%Nb samples at different deformation

| Sample | Deformation (%) | (002) | | (102) | | (103) | |
|---|---|---|---|---|---|---|---|
| | | $\varepsilon_F$ (Å) | $\rho$ (m$^{-2}$)×10$^{-15}$ | $\varepsilon_F$ (Å) | $\rho$ (m$^{-2}$)×10$^{-15}$ | $\varepsilon_F$ (Å) | $\rho$ (m$^{-2}$)×10$^{-15}$ |
| Zircaloy-2 | 30 | 153 | 6.8 | 146 | 7.0 | 144 | 7.8 |
| | 50 | 141 | 7.3 | 133 | 7.9 | 134 | 8.6 |
| | 70 | 126 | 8.9 | 121 | 9.1 | 119 | 9.5 |
| Zr-2.5%Nb | 30 | 132 | 7.2 | 141 | 7.1 | 97 | 8.3 |
| | 50 | 91 | 10.3 | 99 | 11.2 | 76 | 12.2 |
| | 60 | 68 | 15.4 | 69 | 15.3 | 59 | 16.1 |

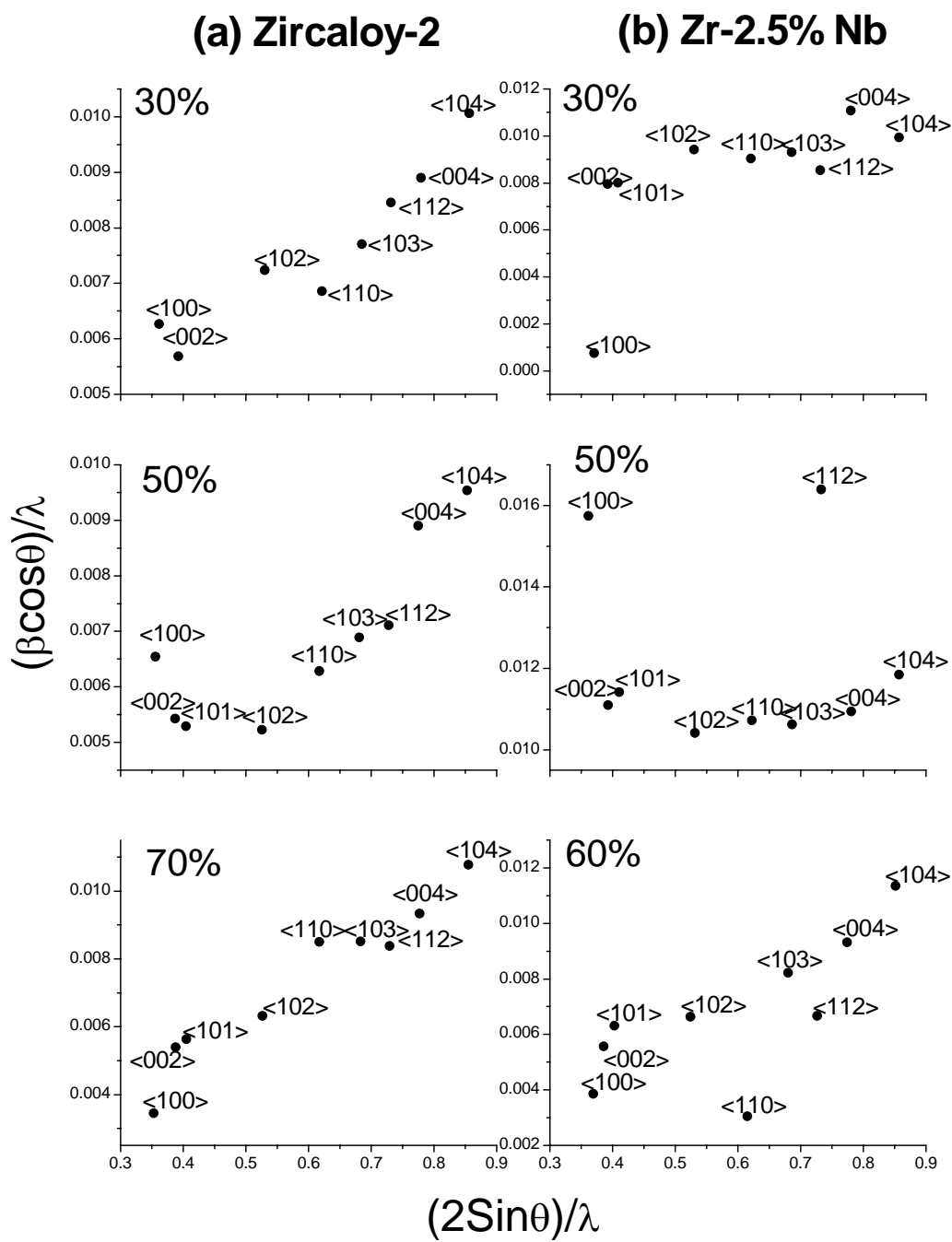

Fig.1. Williamson-Hall plot for (a) Zircaloy-2 and (b) Zr-2.5%Nb at different deformations

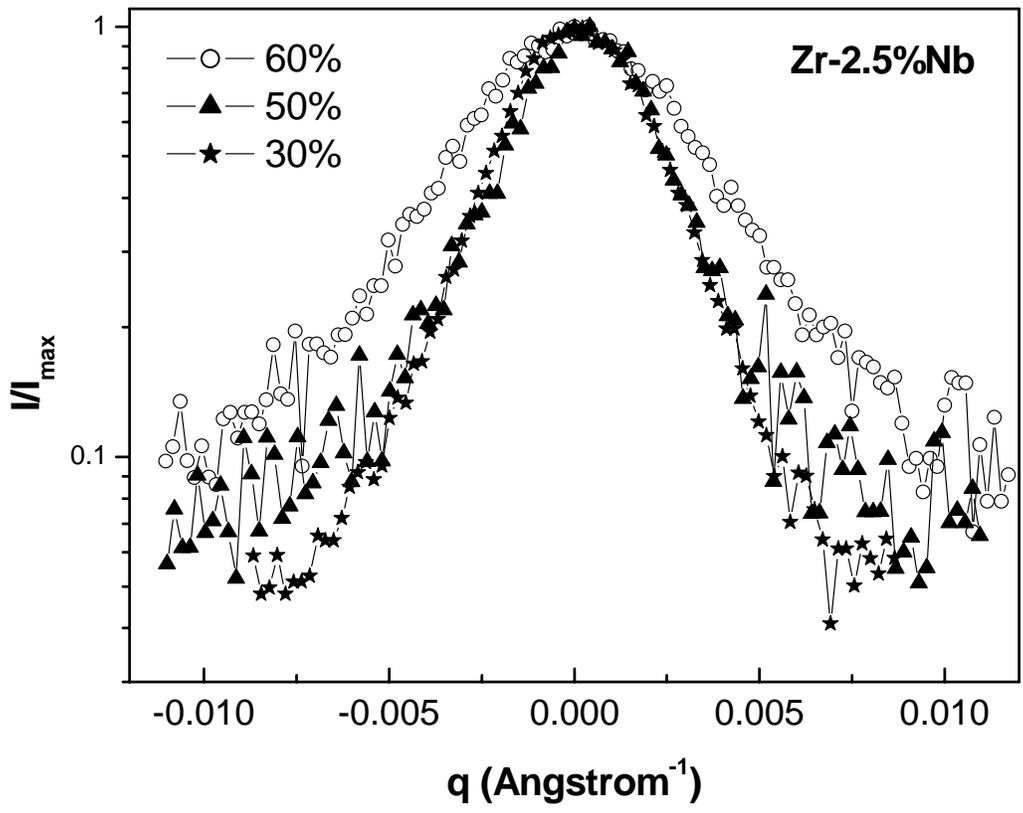

Fig. 2. (002) peaks of the Zr-2.5%Nb sample at different deformations

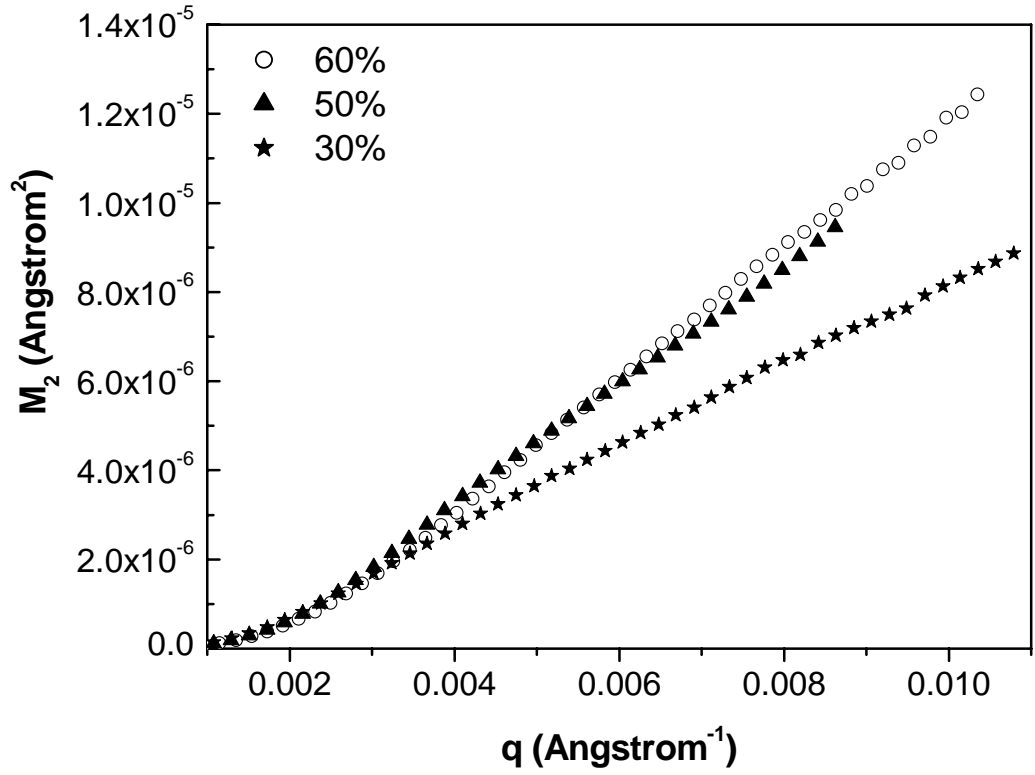

Fig. 3. Second order restricted moment of the (002) peaks of the Zr-2.5%Nb sample at different deformations

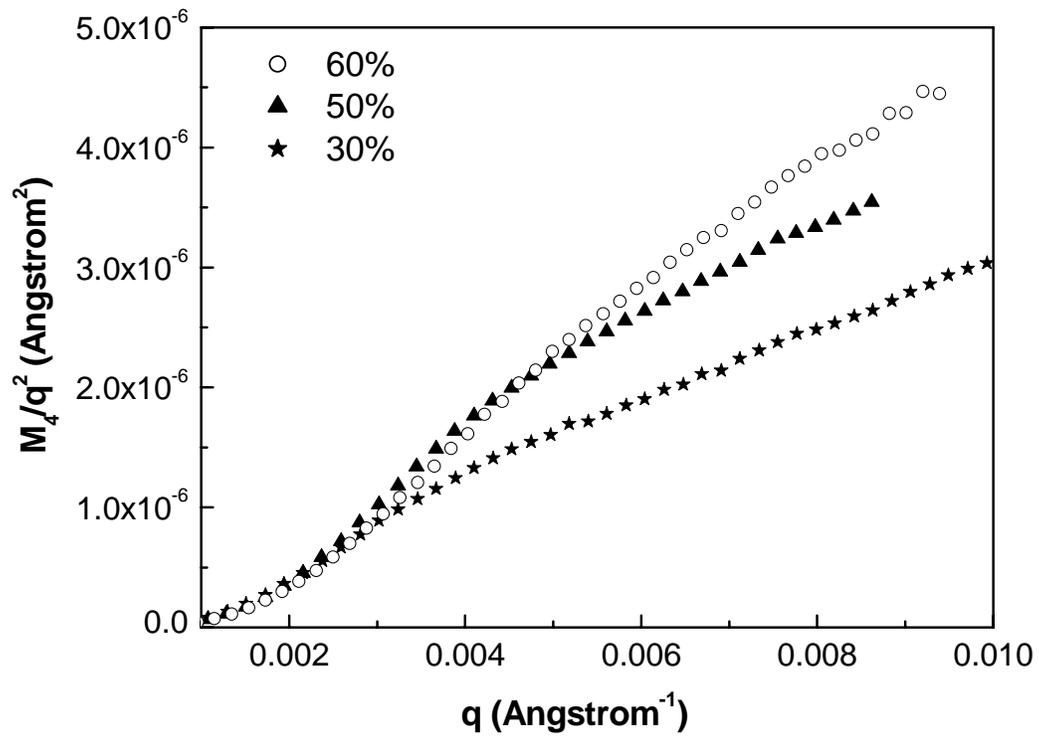

Fig. 4. Fourth order restricted moments of the (002) peaks of the Zr-2.5%Nb sample at different deformations

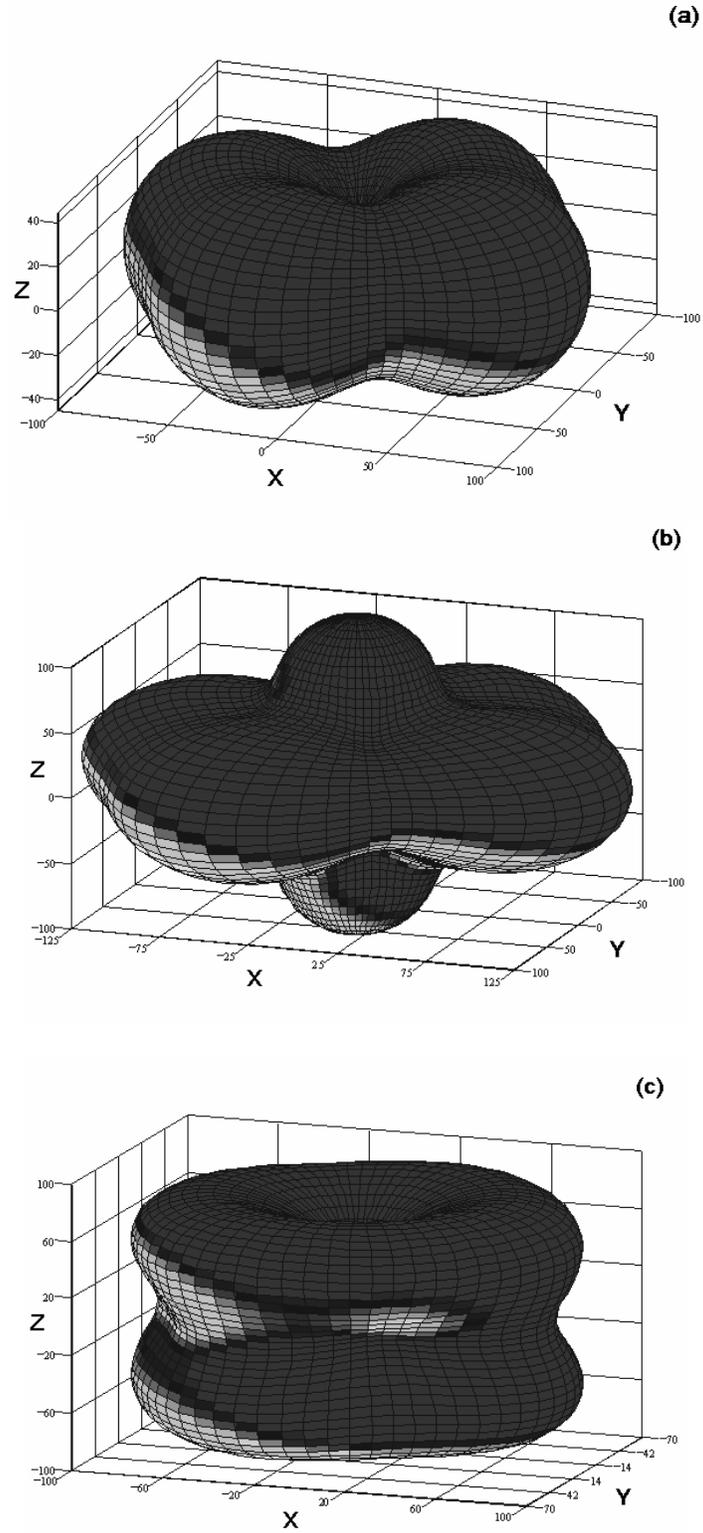

Fig. 5. The three-dimensional anisotropic strain distribution for Zircaloy-2 at different deformations (a) 30%, (b) 50% and (c) 70%. The scale is in $\delta d/d \times 10^{-4}$ strain.

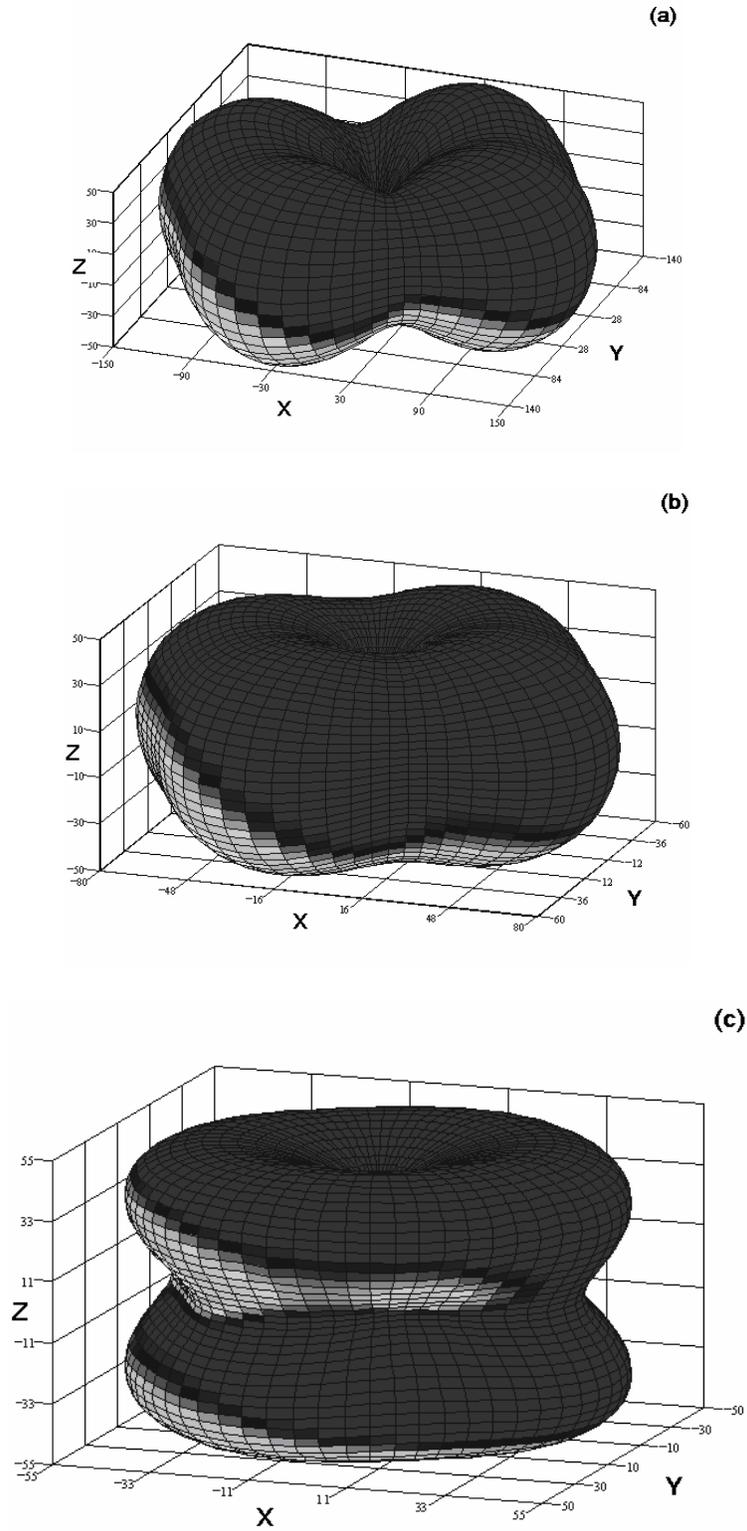

Fig. 6. The three-dimensional anisotropic strain distribution for Zr-2.5%Nb at different deformations (a) 30%, (b) 50% and (c) 60%. The scale is in $\delta d/d \times 10^{-4}$ strain.